\documentclass[letterpaper,twocolumn,showpacs,showkeys,preprintnumbers]{revtex4}
\usepackage{graphicx}

\begin{document}
\title{Softening of Spin-Wave Stiffness near the Ferromagnetic Phase Transition in Diluted Magnetic Semiconductors}
\author{Shih-Jye Sun\footnote{Electronic mail: sjs@nuk.edu.tw}}
\affiliation{Department of Applied Physics, National University of
Kaoshiung, Kaoshiung 811, Taiwan, ROC}
\author{Hsiu-Hau Lin}
\affiliation{Department of Physics, National Tsing-Hua University, Hsinchu 300, Taiwan\\
Physics Division, National Center for Theoretical Science, Hsinchu
300, Taiwan}

\date{\today}
\begin{abstract}
Employing the self-consistent Green's function approach, we
studied the temperature dependence of the spin-wave stiffness in
diluted magnetic semiconductors. Note that the Green's function
approach includes the spatial and temperature fluctuations
simultaneously which was not possible within conventional Weiss
mean-field theory. It is rather interesting that we found the
stiffness becomes dramatically softened as the critical
temperature is approached, which seems to explain the mysterious
sharp drop of magnetization curves in samples within diffusive
regime.
\end{abstract}
\pacs{75.30.Ds, 75.40.Gb, 75.50.Dd} \keywords{magnetic
semiconductor, spintronics, Green's function, softening effect.}
\maketitle

Spintronics\cite{spintronics} brings out an industrial renaissance
in the past decades for its powerful usage of the extra spin
degrees of freedom in many electronic applications. For instances,
giant magneto-resistance (GMR)
  had been applied to read-write head of computer hard disk and tunneling magneto-resistance (TMR)
   to the newest nonvolatile memory for MRAM. The huge success kicked off the intense
   investigations on possible realizations of similar devices in semiconducting materials,
   which can be directly integrated with the existing industrial techniques.
    The idea has enjoyed its primary success in the so-called diluted magnetic semiconductors (DMS),
    where the magnetic ions are doped into the host semiconductors and lead to a ferromagnetic phase.

The key issue at current stage is how to raise the critical
temperature so that the ferromagnetic order is robust even at room
temperature. Taking the well-known material (Ga,Mn)As as
example\cite{GaAs,furdyna1,furdyna2},
 it was demonstrated that the critical temperature can be raised significantly by thermal annealing.
 However, the highest critical temperature at the time of writing is around 160 K\cite{hightc}, which is still far
  from the goal for room-temperature DMS. Based on Zener model, Dietl {\it et al.} proposed to
  look for room-temperature DMS in wide bandgap semiconductors and oxides such as
  GaN, ZnO, TiO$_2$\cite{Matsumoto01,Theodoropoulou01,Lee02}. While the critical temperatures
  in these materials are typically higher compared with (Ga,Mn)As, clustering seems to be a
  serious problem which prevents their potential usage in realistic devices.

While the issue of raising up the critical temperature seems to lie in experimentalists'
hands, we believe a better understanding of the ferromagnetic phase would also help.
After intense theoretical investigations, the origin of ferromagnetism in DMS is
believed to be carrier mediated\cite{rkky}. Typically, the doped transition metal
ions provide the localized impurity spins with small direct exchange (often antiferromagnetic)
among themselves. Furthermore, the experimental results show that only 10$\sim$30$\%$ of the
doped magnetic ions contribute itinerant carriers\cite{doping} into the host semiconducting bands.
Through double-exchange mechanism, the itinerant carriers mediated the indirect exchange
interactions between the localized impurity spins and the ferromagnetic order sets in when
the system is cooled below the critical temperature.

While the origin of the ferromagnetic phase is more or less clear,
some of its physical properties remain puzzling. To achieve a full
understanding, it is crucial to include the exchange coupling
between itinerant and localized spin densities, thermal
fluctuations, the random locations of the doped ions, the
realistic band structure and the repulsive interaction between
itinerant carriers. Since it is almost impossible to incorporate
all effects in single formalism, one needs to glue up piecewise
information from different approaches.

In this paper we employ the self-consistent Green's function
method\cite{PLA}, which includes the spatial and thermal fluctuations simultaneously,
to study the temperature dependence of spin-wave stiffness in DMS. Since our goal is to
demonstrate the peculiar temperature dependence, we simplify the realistic band structure
by the single-band approximation, at price of sacrificing the {\em quantitative} description
for realistic materials. However, within single-band approximation, we were able to establish
the close tie between the softening of stiffness and the sudden drop in magnetization curves\cite{GaAs}.
Moreover, our self-consistent Green's function approach also shows the appearance of concave
magnetization curve\cite{scat} in the regime where itinerant carriers are dilute. This implies
a smooth crossover from the diffusive regime to the localized regime with strong disorder.
Since previous studies including the full six bands only renormalize various physical
parameters and twist the phase boundaries\cite{macdono}, we expect our results to be {\em qualitatively} robust.

The organization of the paper is the following: in the Section II
of this paper, we derive the formalism for the self-consistent
Green's function approach. In Section III, we show our numerical
results and discuss connections to other approaches in the
literature.

The sd model is proper to describe the DMS systems including a
strong exchange interaction between local spins, which come from
electrons in d orbits of transition atom, and itinerant spins
around whole system, which come from impurity doping donation. The
Hamiltonian of the sd model can be expressed by
\begin{equation}
H= H_{0} + J \int d^{3}r S(r) \cdot \sigma(r), \label{kondo},
\end{equation}
the first term $H_{0}$ is the kinetic energy of the itinerant
carriers and the second term is the exchange interaction between
itinerant carrier spins and the localized spin moments, where the
spin density of the localized moments is $S(r) =\sum_{I}
\delta^{3}(r-R_{I}) S_{I}$ and the itinerant spin density is
$\sigma(r) = \psi^{\dag}(r) (\tau/2) \psi(r)$. In momentum
representation the Hamiltonian is expressed as
\begin{eqnarray}
  H&=& H_{k}+H_{J}\nonumber\\
  &=&\sum_{k,\sigma}\varepsilon_{k}c^{\dag}_{k,\sigma}c_{k,\sigma}-c\frac{J}{2}\sum_{k}S^{+}_{k}\sigma^{-}_{k}
        \nonumber\\
  & &-c\frac{J}{2}\sum_{k}S^{-}_{k}\sigma^{+}_{k}-cJ\sum_{k}S^{z}_{k}\sigma^{z}_{-k},
\end{eqnarray}
where the constant c is the density of the magnetic ions in DMS
and $J$ is the magnetic coupling integral in the unit of $eV
nm^3$. Since the RKKY interaction dominates the magnetism in DMS,
 to investigate the magnon dispersion is necessary for studying in microscopic. In order to obtain the magnon dispersion of
local spin in DMS we have to calculate the retarded local spin
Green's function defined by
\begin{eqnarray}
    G_{i,j}(t)&=& \langle\langle S^{+}_{i}(t);S^{-}_{j}(0)y
    \rangle\rangle\nonumber\\
    &=&-i\theta(t)\langle [S^{+}_{i}(t),S^{-}_{j}(0)]
    \rangle,
\end{eqnarray}
where $\theta(t)$ is a step function of time $t$,
$\langle\cdots\rangle$ representing the expectation value and
$[\cdots]$ is the commutator. It is more convenient to derive this
Green's function in momentum space. Through the Fourier
transformation to obtain the spin Green's function in momentum
representation, we get
\begin{equation}\label{sw}
    G(q,t)=\langle\langle S^{+}_{q}(t);S^{-}(0,0)\rangle\rangle.
\end{equation}

Employing the equation of motion to equation (\ref{sw}) makes the
spin Green's function calculation reduced to
\begin{equation}\label{equation}
i\frac{d}{dt}G(q,t)=\varphi+\langle\langle
[S^{+}_{q}(t),H];S^{-}(0,0) \rangle\rangle,
\end{equation}
where function $\varphi$ comes from the derivative of time for
step function.  Following commutation rules in RPA(random phase
approximation) are we used in this paper, there are
\begin{eqnarray}
  [S^{+}_{q},S^{-}_{k}] &=&\frac{2}{N}\sum_{\ell}e^{i(q-k)\cdot R_{\ell}} S^{z}_{\ell}  \nonumber\\
    &=& 2 \langle S^{z} \rangle \delta_{q,k},
\end{eqnarray}

\begin{equation}
[S^{+}_{q},S^{z}_{k}]=-\frac{1}{N}S^{+}_{q+k},
\end{equation}

\begin{equation}
    [\sigma^{+}_{k},\sigma^{-}_{k^{'}}]=\sum_{p}(c^{\dag}_{p+k,\uparrow}c_{p-k^{'},\uparrow}-c^{\dag}_{p+k^{'}+k,\downarrow}
    c_{p,\downarrow})
\end{equation}

and
\begin{equation}
   [\sigma^{+}_{k},\sigma^{z}_{k^{'}}]=-\frac{1}{N}\sigma^{+}_{k+k^{'}},
\end{equation}
where $\langle S^{z} \rangle$ is the expectation value of
magnetization for local spin. Employing these commutation rules to
equation (\ref{equation}) results to the equation
\begin{eqnarray}\label{gen}
   \omega G(q,\omega)&=& \varphi-cJ\langle S^{z} \rangle  \langle\langle \sigma^{+}_{q};S^{-}(0,0)\rangle\rangle
   \nonumber\\
   & &+c J \langle\sigma^{z}\rangle\langle\langle S^{+}_{q};S^{-}(0,0)
   \rangle\rangle,
\end{eqnarray}

where $\langle \sigma^{z}\rangle$ is the expectation value of the
magnetization for itinerant carriers. We found the derivation
process for local spin Green's function resulting to another new
Green's function $\zeta(k,\omega)=\langle\langle
   \sigma^{+}_{k};S^{-}(0,0)\rangle\rangle$ in the meanwhile. It
   is clearly the new resulted Green's function reveals the physics
   that the exchange interaction between local spins needs the itinerant
   carriers' mediation. Therefore we need to
   calculate the new Green's function by the same way,

\begin{equation}\label{fg}
   i\frac{d}{dt}\langle\langle \sigma^{+}_{q};S^{-}(0,0)
   \rangle\rangle=\langle\langle [\sigma^{+}_{q},H];
   S^{-}(0,0)\rangle\rangle.
\end{equation}
The commutator in equation (\ref{fg}) results to
\begin{equation}
    [\sigma^{+}_{q},H_{k}]=\sum_{p}(\varepsilon_{p}-\varepsilon_{p+q})c^{\dagger}_{p+q,\uparrow}c_{p,\downarrow}
\end{equation}
\begin{equation}
    [\sigma^{+}_{q},H_{J}]=-\frac{c J}{2}\sum_{k}[\sigma^{+}_{q},\sigma^{-}_{k}]S^{+}_{k}-c J\sum_k
    [\sigma^{+}_{q},\sigma^{z}_{k}]S^{z}_{k},
\end{equation}
where $\varepsilon_{k}$ is the electrons kinetic energy with
momentum $k$. Apply the Fourier transformation and RPA for
carriers density to the new Green's function $\zeta$ resulting to

\begin{eqnarray}\label{iti}
  \omega\zeta(q,\omega)&=&\sum_{p}(\varepsilon_{p}-\varepsilon_{p+q})\langle\langle c^{\dag}_{p+q,\uparrow}c_{p,\downarrow};S^{-}(0,0)
  \rangle\rangle\nonumber\\
  & &-\frac{c J}{2}\sum_{p}(\langle c^{\dag}_{p+q,\uparrow}c_{p+q,\uparrow}\rangle-\langle c^{\dag}_{p,\downarrow}c_{p,\downarrow}\rangle) \nonumber\\
  & & \times G(q,\omega)+c J \langle S^{z} \rangle \zeta(q,\omega).
\end{eqnarray}
Abstracting momentum $p$ from $\zeta(q,\omega)$ in equation
(\ref{iti}) results to a relation
\begin{eqnarray}\label{ifi}
    \langle\langle c^{\dagger}_{p+q,\uparrow}c_{p,\downarrow};S^{-}(0,0)
    \rangle\rangle &=& \frac{c J}{2}\frac{\langle c^{\dagger}_{p,\downarrow}c_{p,\downarrow}\rangle-\langle c^{\dagger}_{p+q,\uparrow}c_{p+q,\uparrow}\rangle}{\omega-\varepsilon_{p}+\varepsilon_{p+q}-c J\langle S^{z} \rangle}
    \nonumber\\
    & & \times G(q,\omega).
\end{eqnarray}

Combining equations (\ref{gen}) and (\ref{iti}) results to a
closed form of Green's function equation $G(q,\omega)$,
\begin{eqnarray}\label{ggen}
   (\omega-c J\langle \sigma^z \rangle &+& \frac{c J^2}{2}\langle S^z
   \rangle\sum_{p}\frac{\langle c^{\dagger}_{p,\downarrow}c_{p,\downarrow}\rangle-\langle c^{\dagger}_{p+q,\uparrow}c_{p+q,\uparrow}\rangle}{\omega-\varepsilon_{p}+\varepsilon_{p+q}-c J\langle S^{z}
   \rangle})\nonumber\\
   & & \times G(q,\omega)=\varphi,
\end{eqnarray}
where the $\langle c^{\dagger}_{p,\sigma}c_{p,\sigma} \rangle =
f_{p,\sigma}=(\beta \varepsilon_{p,\sigma}+1)^{-1}$ is the carrier
density for spin $\sigma$ and $\beta=1/K_BT$. The poles in
eq.(\ref{ggen}) represent magnon excitations. In the diluted
limitation, the minority is the itinerant carriers which could be
seen as a free carrier gas with effective mass
$m^{*}=0.5m_e$\cite{rkky}, where $m_e$ is the free electron mass,
and in the ferromagnetic state the majority of local magnetic ions
produce an effective magnetic field causing both spins to split
with Zeeman energy $c J\langle S^{z}\rangle$. The kinetic energies
of different spins are $\varepsilon_{k,\sigma}=\hbar^2
k^2/2m^{*}\mp \sigma 1/2 cJ\langle S^z \rangle$ respectively.
Therefore the magnon excitation energy for each momentum $q$ is

\begin{eqnarray}
  \nonumber \omega_q &=& J \langle{\sigma^z}\rangle -\frac{c J^2}{8\pi^2} \langle{S^z}\rangle \times \{\int^{\infty}_{0} \frac{km^{\ast}f_{k\uparrow}}{\hbar^2q} \nonumber\\
   & & \times \ln(\frac{\omega_q-c J\langle {S^z}\rangle - \frac{\hbar^2q^2}{2m^\ast}-\frac{\hbar^2kq}{m^\ast}}
  {\omega_q-c J\langle {S^z}\rangle - \frac{\hbar^2q^2}{2m^\ast}+\frac{\hbar^2kq}{m^\ast}})dk + \nonumber\\
  & & \int^{\infty}_{0} \frac{km^{\ast}f_{k\downarrow}}{\hbar^2q}
  \ln(\frac{\omega_q-c J\langle {S^z}\rangle +\frac{\hbar^2q^2}{2m^\ast}-\frac{\hbar^2kq}{m^\ast}}
  {\omega_q-c J\langle {S^z}\rangle +\frac{\hbar^2q^2}{2m^\ast}+\frac{\hbar^2kq}{m^\ast}})\nonumber\\
  & & \times dk\}.
\end{eqnarray}

Finally, we could utilize the Callen's arbitrary spin formula,
\begin{eqnarray}\label{callen}
    \langle S^{z}  \rangle &=& \frac{[S-\Phi(S)][1+\Phi(S)]^{2S+1}+[S+1+\Phi(S)]\times
    }{[1+\Phi(S)]^{2S+1}-[\Phi(S)]^{2S+1}}\nonumber\\
    & &\frac{[\Phi(S)]^{2S+1}}{}
\end{eqnarray}
to obtain the magnetization values, where
$\Phi(S)=\frac{1}{N}\sum_{q}(\exp(\beta \omega_q)-1)^{-1}$ is the
magnon number.

The anomalous temperature dependance of magnetization of sharp
drop in the vicinity of $T_C$ represents in our theoretical result
as shown in the insert of Fig. 1. is consistent with experimental
result implying an important interaction existing in the
magnetization collapse region. According to the RKKY mechanism the
magnetism is established by itinerant carriers mediation. The
sharp drop of magnetization reveals a possibility of
carrier-magnon decoupling in the vicinity of $T_C$. This
decoupling effect reduces the effective magnetic interaction and
the magnetization disappears in decoupling completely.

\begin{figure}[t]
\centering
\includegraphics[width=8.5cm]{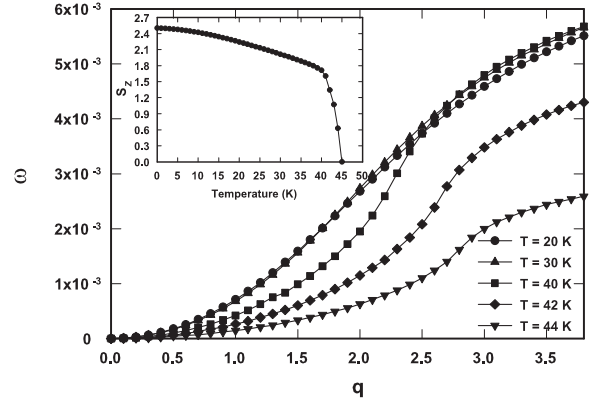}
\caption{We take the exchange coupling and the effective mass are
fixed at typical values $J = 0.15$ eV nm$^3$, $m^* = 0.5 m_e$ and
the ratio of itinerant and localized spin densities fixed at
$c^*/c$ =0.1 to calculation resulting to $T_C=45K$. In the insert
of the figure shows a sharp drop of magnetization in the vicinity
of $T_C$. In the sharp drop region the magnon shows a softening
effect.}
\end{figure}

The dispersion of spin wave derived from the conventional spin
wave theories\cite{spinwave} is temperature independent, which is
an intrinsic characteristic for many kinds of magnets. The Fig. 1
exhibits the dispersion of magnon from our theoretical calculation
showing a temperature independent dispersion at far from $T_Cs$,
which reveals a result hat the temperature independence of normal
spin wave existing at robust magnetism region, meanwhile it
reveals an obvious magnon softening effect in the vicinity of
$T_C$ leading to the magnetization falling down sharply.
Interestingly, this softening effect starts from small magnon
momentum $qs$ then extending to whole dispersion region
eventually. From the conventional spin wave theory as the $q \ll$,
the magnon dispersion relation has $\omega(q)=D q^{2}$, where the
stiffness constant $D \propto J^{'}$ and $J^{'}$ is the magnetic
coupling integral between two separated spins. From the linear
response theory we have derived\cite{APL} before, the coupling
$J^{'}$ is $\propto J^{2}$, where the $J$ is the coupling between
itinerant spins and local spins. Therefore this softening effect
results to $D$ decreasing, which reveals a fact that the effective
coupling $J$ reduces and gives an implication with carrier-magnon
decouple in system.

In the conclusion, by our theoretical study the anomalous
magnetization sharp drop in the vicinity of $T_C$ comes from the
magnon softening effect, and this softening effect possibly comes
from the carrier-magnon decoupling. We thanks the support of National Center of Theoretical Science of
Taiwan.


\begin{thebibliography}{99}
\bibitem{scat}
From our present calculation shows adding an impurity level to
hybridize the conduction band results to an obvious concave
magnetization cure.
\end{thebibliography}
\end{document}